\documentclass[aps, twocolumn, superscriptaddress, floatfix, nofootinbib, amsmath, amssymb, preprintnumbers]{revtex4-1}

\usepackage{dcolumn}% Align table columns on decimal point
\usepackage{verbatim}

\usepackage{breakurl}

\usepackage{lmodern}

\pdfoutput=1
\usepackage{graphicx}
\usepackage{bm}
\usepackage{amsmath}
\usepackage{amsfonts}
\usepackage{amssymb}
\usepackage{multirow}
\usepackage[colorlinks,linktocpage,linkcolor=cyan,citecolor=cyan]{hyperref}
\usepackage{adjustbox}
\usepackage{array}
\usepackage[capitalise]{cleveref}
\usepackage{acro}
\usepackage[utf8]{inputenc}
\usepackage{CJKutf8}
\usepackage{amssymb}

\usepackage{tikz}
\usepackage{pgfplots}
\pgfplotsset{compat=newest,every axis plot/.append style={line width=1pt}}

\crefname{figure}{Fig.}{Figs.}
\Crefname{figure}{Fig.}{Figs.}
\def\({\left(}
\def\){\right)}
\def\[{\left[}
\def\]{\right]}

\newcommand{\be}{{\begin{eqnarray}}}
\newcommand{\ee}{{\end{eqnarray}}}

\newcommand{\Gpcyr}{{\rm Gpc^{-3}~yr^{-1}}}
\newcommand{\overbar}[1]{\mkern 1.5mu\overline{\mkern-1.5mu#1\mkern-1.5mu}\mkern 1.5mu}

\DeclareAcronym{LIGO}{
  short =LIGO ,
  long = Laser Interferometer Gravitational-Wave Observatory ,
  short-plural = ,
}
\DeclareAcronym{LVK}{
  short = LVK ,
  long = {Advanced LIGO, Virgo, and KAGRA Collaborations} ,
  short-plural = ,
}
\DeclareAcronym{GW}{
  short = GW ,
  long = gravitational wave ,
  short-plural = s ,
}
\DeclareAcronym{SGWB}{
  short = SGWB ,
  long = stochastic gravitational-wave background ,
  short-plural = s ,
}
\DeclareAcronym{CBC}{
  short = CBC ,
  long = compact binary coalescence ,
  short-plural = s ,
}
\DeclareAcronym{BH}{
  short = BH ,
  long = black hole ,
  short-plural = s ,
}
\DeclareAcronym{BBH}{
  short = BBH ,
  long = binary black hole ,
  short-plural = s ,
}
\DeclareAcronym{PBH}{
  short = PBH ,
  long = primordial black hole ,
  short-plural = s ,
}
\DeclareAcronym{SNR}{
  short = SNR ,
  long = signal-to-noise ratio ,
  short-plural = s ,
}
\DeclareAcronym{IMRPPv2}{
  short = ,
  long = {\normalsize IMRP}{\footnotesize HENOM}{\normalsize P}v2 ,
  short-plural = ,
}
\DeclareAcronym{PTA}{
  short = PTA ,
  long = pulsar timing array ,
  short-plural = s ,
}
\DeclareAcronym{SFR}{
  short = SFR ,
  long = star formation rate ,
  short-plural =  ,
}
\DeclareAcronym{FRW}{
  short = FRW ,
  long = Friedman-Robertson-Walker ,
  short-plural =  ,
}
\DeclareAcronym{IMR}{
  short = IMR ,
  long = inspiral-merger-ringdown ,
  short-plural =  ,
}
\DeclareAcronym{LISA}{
	short = LISA ,
	long  = Laser Interferometer Space Antenna,
  short-plural =  ,
}
\DeclareAcronym{ET}{
	short = ET ,
	long  = Einstein Telescope,
  short-plural =  ,
}
\DeclareAcronym{CE}{
	short = CE ,
	long  = Cosmic Explorer,
  short-plural =  ,
}
\DeclareAcronym{BBO}{
	short = BBO ,
	long  = big bang observer,
  short-plural =  ,
}
\DeclareAcronym{DECIGO}{
	short = DECIGO ,
	long  = Deci-hertz Interferometer Gravitational wave Observatory,
  short-plural =  ,
}
\DeclareAcronym{ABH}{
	short = ABH ,
	long  = astrophysical black hole,
  short-plural = s ,
}

\begin{document}

\title{Probing primordial black holes with anisotropies in stochastic gravitational-wave background}

\author{Sai Wang}
\email{Corresponding author: wangsai@ihep.ac.cn}
\affiliation{Theoretical Physics Division, Institute of High Energy Physics, Chinese Academy of Sciences, Beijing 100049, People's Republic of China}
\affiliation{School of Physical Sciences, University of Chinese Academy of Sciences, Beijing 100049, People's Republic of China}

\author{Valeri Vardanyan}
\email{Corresponding author: valeri.vardanyan@ipmu.jp}
\affiliation{Kavli Institute for the Physics and Mathematics of the Universe (WPI), UTIAS, The University of Tokyo, Kashiwa, Chiba 277-8583, Japan}

\author{Kazunori Kohri}
\email{Corresponding author: kohri@post.kek.jp}
\affiliation{Theory Center, IPNS, KEK, 1-1 Oho, Tsukuba, Ibaraki 305-0801, Japan}
\affiliation{International Center for Quantum-field Measurement Systems for Studies of the Universe
and Particles (QUP, WPI), KEK, 1-1 Oho, Tsukuba, Ibaraki 305-0801, Japan}
\affiliation{The Graduate University for Advanced Studies (SOKENDAI), 1-1 Oho, Tsukuba, Ibaraki 305-0801, Japan}
\affiliation{Kavli Institute for the Physics and Mathematics of the Universe (WPI), UTIAS, The University of Tokyo, Kashiwa, Chiba 277-8583, Japan}

\begin{abstract} 
Primordial black holes, if considered to constitute a significant fraction of cold dark matter, trace the inhomogeneous large-scale structure of the Universe. Consequently, the stochastic gravitational-wave background, originating from incoherent superposition of unresolved signals emitted by primordial black hole binaries, is expected to display anisotropies across the sky. In this work, we investigate the angular correlations of such anisotropies for the first time and demonstrate their difference from the analogous signal produced by astrophysical black hole binaries. We carefully evaluate the associated uncertainties due to shot-noise and cosmic variance, and demonstrate that the studied signal in the low-frequency regime can be differentiated from the signal of astrophysical origin. Our results are particularly promising in the stellar mass-range, where the identification of the merger origin has been particularly challenging.
\end{abstract}

\maketitle

\acresetall
%%%%%%%%%%%%%%%%%%%%%%%%%%%%%%%%%%%%%%%%%%%%%%%

\section{Introduction}\label{sec:introduction}

Observations of \acp{GW} sourced by binary black holes (BBH) \cite{Abbott:2016blz} have stimulated extensive studies on \acp{PBH} (for a review, see Ref. \cite{Sasaki:2018dmp}). 
\acp{PBH} could have been produced in the early stages of the Universe by gravitational collapse of primordial density perturbations, immediately after these have reentered into the Hubble horizon \cite{Hawking:1971ei,Carr:1974nx,GarciaBellido:1996qt,Clesse:2015wea,Dolgov:2013lba,Harada:2013epa,Harada:2016mhb,Khlopov:2008qy,Belotsky:2014kca,Ketov:2019mfc,Zhou:2020kkf}. 
The relative abundance of \acp{PBH} with respect to cold dark matter has been tightly constrained by a variety of astronomical observations (for reviews, see Refs. \cite{Carr:2020gox,Carr:2020xqk}). It has been shown that even a relatively low abundance of \acp{PBH} in the mass-range of current interferometers is capable of accounting for the observed local merger rates of black hole binaries \cite{Nishikawa:2017chy,Sasaki:2016jop}. Note, however, that the observed neutron--star black hole binaries \cite{LIGOScientific:2021qlt} are predominantly of astrophysical origin \cite{Sasaki:2021iuc}, although speculations of the reported neutron stars, alongside the primary components of the binaries, being primordial black holes have also been considered \cite{Wang:2021iwp}. 
The \ac{LVK} \cite{Abbott:2018oah,Authors:2019qbw} have not yet detected compact objects in the subsolar mass-range, which would have been considered to be a smoking gun for the \ac{PBH} scenario. 
There is also no evidence for mergers composed of stellar- and subsolar-mass black holes \cite{Nitz:2020bdb}. As a result, one of the key features of the \ac{PBH} scenario is the distinctive redshift distribution of merger rate at high redshifts $z \gtrsim 10$ which can be probed by future generations of gravitational wave detectors \cite{Nakamura:2016hna,Koushiappas:2017kqm,Chen:2019irf}.

An alternative observable, the \ac{SGWB} \cite{Regimbau:2011rp} produced by the incoherent superposition of \aclp{GW} from all the unresolved \ac{PBH} binaries in the Universe, has been proposed to independently constrain the abundance of \acp{PBH} \cite{Wang:2016ana,Mandic:2016lcn,Clesse:2016ajp,Raidal:2017mfl}. 
In fact, strong upper bounds on the abundance of \acp{PBH} have been obtained \cite{Wang:2016ana,Kapadia:2020pnr} using the null-detection of the \ac{SGWB} by the \ac{LVK} network \cite{TheLIGOScientific:2016dpb,LIGOScientific:2019vic}. A variety of future observations are expected to further improve these constraints \cite{Wang:2019kaf}. 
The \ac{SGWB}s arising in the stellar-mass \ac{PBH} scenario and in astrophysical context are effectively indistinguishable from each other at the current detector sensitivities \cite{Mukherjee:2021ags}.

%\textbf{Move this to a better location: 
%The energy density spectrum of \ac{SGWB} was originally studied in the background spacetime \cite{Allen:1997ad}.}

On top of the directional average, the background also features potentially observable anisotropies which can provide additional useful information. The anisotropies of the \ac{SGWB} originate from the spatial clustering of \ac{GW} sources\footnote{In this work we do not consider the resolved \ac{GW} sources. Nonetheless, the clustering signal of the latter is a sensitive probe for identifying the origin of GW binaries; see Refs.~\cite{Raccanelli:2016cud,Scelfo:2018sny,Mukherjee:2020hyn,Canas-Herrera:2021qxs}.}, which trace the spatial distribution of dark matter \cite{Matsubara:2019qzv,Trashorras:2020mwn,Atal:2020igj,Ding:2019tjk,Belotsky:2018wph}. As we will see, this new window holds the potential of distinguishing the \ac{PBH} scenario from the \ac{ABH} one. In this paper we provide the theoretical modelling of the spectra characterizing the angular correlations of anisotropies. We, particularly, for the first time, present the computation of the angular power spectra of the \ac{SGWB} anisotropies in the context of \acp{PBH}. We also reproduce the computation in the astrophysical scenario, taking into account the Pop--II and Pop--III stellar populations.

In this paper we demonstrate that the angular power spectrum of the \ac{SGWB} provides a complementary pathway towards identifying the origin of black hole binaries detected by gravitational wave detectors. The difference in angular correlations of \ac{PBH} and \ac{ABH} scenarios relies in the differing spatial clustering properties of the two \ac{BBH} populations (see e.g. \cite{Ali-Haimoud:2017rtz}), as well as the redshift dependence of the merger rates. The latter is a monotonically increasing function of redshift in the \ac{PBH} scenario \cite{Sasaki:2016jop,Raidal:2017mfl,Ali-Haimoud:2017rtz,Chen:2018czv}, and traces the \ac{SFR} in the \ac{ABH} scenario \cite{deSouza:2011ea,Dominik:2012kk,Vangioni:2014axa,Kinugawa:2014zha}. 

The rest of the paper is organized as follows. 
In Sec.~\ref{sec:rate} we briefly review the merger rates of \acp{BBH} of primordial and astrophysical origins. 
In Sec.~\ref{sec:density} we summarize the formalism used for computing the angular correlations of the anisotropic \ac{SGWB}. 
We present our results in Sec.~\ref{sec:result} and conclusions in Sec.~\ref{sec:conc}.

\section{Merger rate of black hole binaries}\label{sec:rate}

There are two widely-considered channels for \ac{PBH} binary formation. In the early Universe channel \ac{PBH} binaries form due to torque exerted by all the neighboring \acp{PBH} as well as the linear density perturbations \cite{Sasaki:2016jop,Raidal:2017mfl,Ali-Haimoud:2017rtz,Chen:2018czv}. In the late Universe one, instead, binaries are formed due to close encounters of \acp{PBH} in dark matter halos \cite{Nishikawa:2017chy,Bird:2016dcv,Raidal:2017mfl}. In order to match the LIGO-Virgo local merger rate, the abundance of \ac{PBH} should be less than $\mathcal{O}(10^{-3})$ in the early Universe channel \cite{Sasaki:2016jop,Ali-Haimoud:2017rtz}, while an $\mathcal{O}(1)$ fraction is required in the late Universe channel \cite{Bird:2016dcv}. Assuming no significant disruption of early Universe binaries, late Universe binaries would only constitute a negligible fraction of the total binaries and they can be safely neglected in our analysis. For simplicity we assume a monochromatic mass distribution of \acp{BH}, with the component masses given by $m_0=23M_\odot$. Changing the \ac{PBH} mass to other values would not change our predictions significantly. We have chosen this relatively large mass in order to facilitate comparisons with the Pop-II and Pop-III sources. We would like to stress, however, that the formalism can be easily generalized to other mass distributions. While narrow-shaped mass distributions, similar to those considered in Ref.~\cite{Chen:2018czv}, would not substantially change our results, broad distributions should be studied more carefully in a future work.

%As a function of redshift $z$, the comoving merger rates $\mathcal{R}(t(z))$ in PBH (red curve) and ABH (blue curve) scenarios are shown in Fig.~\ref{fig:0}.

The comoving merger rate of \ac{PBH} binaries in $\Gpcyr$ units, evaluated at cosmic time $t$, is given by \cite{Chen:2018czv}
\begin{align}\label{eq:pr}
\mathcal{R}_{\mathrm{PBH}} = A &\left(\frac{t_0}{t}\right)^{\frac{34}{37}} \frac{f^{2}}{ (f^{2}+\sigma_{\mathrm{eq}}^{2})^{\frac{21}{74}}}\times\nonumber\\ &\left(\frac{m}{M_\odot}\right)^{-\frac{32}{37}}\delta\left(\frac{m}{M_\odot}-\frac{m_0}{M_\odot}\right) \ ,
%\mathcal{R}_{\mathrm{PBH}}(z)=A_{\mathrm{PBH}}\left(\frac{t(z)}{\tau}\right)^{-\frac{34}{37}}\ ,
\end{align}
where $A\simeq3.8\times10^{6}$ is a constant amplitude, $m$ denotes the \ac{PBH} mass, $t_0$ is the present age of the Universe, $f_{\mathrm{PBH}}$ is the fraction of dark matter in the form of \acp{PBH}, and $\sigma_{\mathrm{eq}} \simeq 0.005$ is the variance of overdensities of the rest of dark matter on scales of order $\mathcal{O}(10^{-2}-10^5)M_\odot$ at the epoch of matter-radiation equality \cite{Ali-Haimoud:2017rtz}. 
We have assumed that the primordial curvature perturbations producing the \acp{PBH} are almost Gaussian, and the initial \ac{PBH} clustering can be safely neglected; see e.g. Refs.\cite{Suyama:2019cst,Matsubara:2019qzv}. Effects of initial clustering will be explored in the future. In Eq.~(\ref{eq:pr}), we have additionally disregarded the effects of binary disruptions \cite{Raidal:2018bbj,Hutsi:2020sol}, which is justified when $f_{\mathrm{PBH}}\simeq10^{-3}$ and when initial clustering is neglected. 

\begin{figure}
    \includegraphics[width =1. \columnwidth]{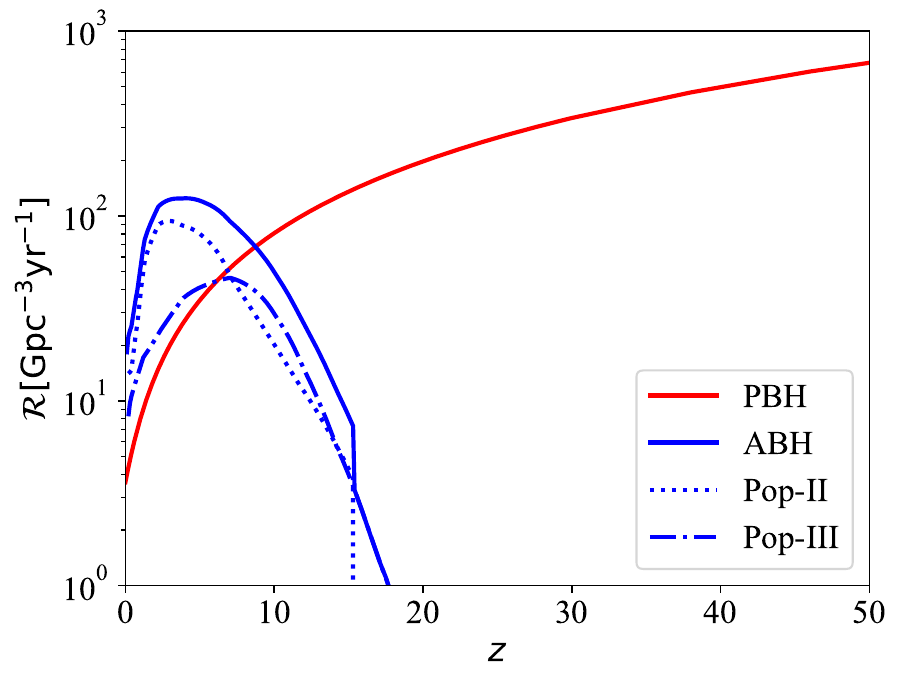}
    \caption{Comoving merger rate $\mathcal{R}(t(z))$ as a function of redshift $z$, in PBH (red) and ABH (blue) scenarios. For comparison, we plot the comoving merger rates of Pop-II (dotted) and Pop-III (dash-dotted) binaries separately. }\label{fig:0}
\end{figure}

The existing observational constraints suggest $f_{\mathrm{PBH}}\simeq 10^{-3}$ for $m_{\mathrm{PBH}}=23M_\odot$ at the 90\% confidence upper limit; see e.g., Ref.~\cite{Carr:2020gox} for a recent summary. Equation (\ref{eq:pr}) implies that the merger rate monotonically decreases with $t$, or equivalently, increases with redshift $z$ (see the red curve in Fig.~\ref{fig:0}). Throughout this paper we will fix all of the cosmological parameters to their best-fit values inferred by the Planck 2018 results \cite{Aghanim:2018eyx}.

In contrast to \ac{PBH} binaries, the abundance of \acs{ABH} binaries is closely related to the star-formation processes \cite{deSouza:2011ea,Dominik:2012kk,Vangioni:2014axa,Kinugawa:2014zha}. We take into account the Pop--II and Pop--III sources when estimating the abundance of \ac{ABH} binaries. We adopt the redshift dependence of the event rate as described in Fig.~10 of Ref.~\cite{Nakamura:2016hna}, which is shown as the blue solid curve in Fig.~\ref{fig:0}. This rate arises from a superposition of Pop--II (blue dotted curve) and Pop--III (blue dash-dotted curve) binaries, which are also plotted in Fig.~\ref{fig:0} for comparison. The comoving merger rate of \acp{ABH} is peaked at $z\lesssim 10$ and rapidly decreases at higher redshifts. This crucial difference from the \ac{PBH} scenario has been discussed in context of future \ac{GW} detectors, such as the \ac{DECIGO} and the \ac{BBO} \cite{Nakamura:2016hna}, as well as \ac{ET} and \ac{CE} \cite{Koushiappas:2017kqm,Chen:2019irf}. 
%Please refer to Fig.~10, from which I take data points and make rescaling for the evaluation in next section. 
As in the case of the \ac{PBH} scenario, we assume a monochromatic mass function for the astrophysical black holes as well. This simplifying step allows for more straightforward comparisons of the \ac{PBH} and \ac{ABH} scenarios. It is important to note that the details of the mass distribution is largely uncertain for both of the scenarios. 
There are more recent works on this topic, that also provide the \ac{ABH} merger rate, see e.g., Ref.~\cite{Ng:2020qpk}. However, within large uncertainties, the related theoretical predictions are expected to be overall similar to that of the assumed merger rate in our paper. 
%The merger rate of \acp{ABH} can be expressed as a convolution of \ac{SFR} $R_{\ast}(t_{\mathrm{f}})$ and time delay distribution $P(t_\mathrm{d})$, i.e., \cite{TheLIGOScientific:2016wyq}
%\begin{align}
%\label{eq:ar}
%\mathcal{R}_{\mathrm{ABH}}(z)=A_{\mathrm{ABH}}\int_{50\mathrm{Myr}}^{t_{\mathrm{c}}(z)} R_{\ast}(t_{\mathrm{f}}) P(t_{\mathrm{d}})\frac{1+z}{1+z_{\mathrm{f}}}dt_{\mathrm{d}}\ ,
%\end{align}
%where $t_{\mathrm{d}}$ is the delay time between the formation and the coalescence of binary, $t_{\mathrm{c}}(z)$ is the cosmic time at coalescence corresponding to redshift $z$, $z_{\mathrm{f}}$ is the redshift at the formation time $t_{\mathrm{f}}=t_{\mathrm{c}}-t_{\mathrm{d}}$. 
%We employ the \ac{SFR} model in Ref.~\cite{Vangioni:2014axa} and the time delay model $P(t_{\mathrm{d}})\propto 1/t_{\mathrm{d}}$.  
%Similar to $A_{\mathrm{PBH}}$, the amplitude $A_{\mathrm{ABH}}$ is also determined by calibrating $\mathcal{R}_{\mathrm{ABH}}(z=0)$ to the observed limit \cite{Abbott:2020gyp}. 
%Since \ac{SFR} has a peak at $z\simeq2$, the merger rate $\mathcal{R}_{\mathrm{ABH}}$ is peaked at $z\lesssim1-2$ and decreases monotonically with redshift above the peak redshift. 

\section{Angular power spectrum of the SGWB anisotropies}\label{sec:density}

For the modeling of anisotropies we mostly follow Refs.~\cite{Cusin:2017fwz,Jenkins:2018uac,Contaldi:2016koz} and evaluate the line-of-sight distribution of the \ac{SGWB} as a function of direction on the sky. The projected intensity maps reflect the spatial clustering properties of the \ac{GW} sources, as well as the propagation effects due to inhomogeneous large-scale structure of dark matter. We particularly consider perturbations around spatially-flat \acl{FRW} metric as $ds^2=a^2[-(1+2\phi)d\eta^2+(1-2\psi)\delta_{ij}dx^idx^j]$, where $\psi$ and $\phi$ denote the two Bardeen potentials (assumed to be identical in this work), $a= a(\eta)$ is the scale factor, and $\eta=\int dt/a$ is the conformal time.

We model the directional dependence of the projected \ac{GW} intensity per unit solid angle as 
\begin{align}\label{eq:omegagw}
\Omega(\nu,\mathbf{e})=\frac{1}{\rho_{\mathrm{c}}}\frac{d^{3}\rho(\nu,\mathbf{e})}{d\ln\nu d^{2}\mathbf{e}}
%= \frac{\bar{\Omega}(\nu)}{4\pi}\left(1+\delta(\nu,\mathbf{e})\right)\ ,
= \frac{\overbar{\Omega}(\nu)}{4\pi}+\delta\Omega(\nu,\mathbf{e})\ ,
\end{align}
where $\rho$ is the energy density at an observed frequency $\nu$, $\mathbf{e}$ is a unit vector along the line-of-sight. The critical energy density of the Universe at the present epoch is defined as $\rho_{\mathrm{c}}=3H_{0}^{2}/(8\pi G)$, where $G$ and $H_{0}$ are the gravitational and Hubble constants, respectively. 
Here, $\overbar{\Omega}$ is the homogeneous and isotropic component described previously in \cite{Allen:1997ad,Wang:2016ana,Mandic:2016lcn,Clesse:2016ajp,Raidal:2017mfl}, while $\delta\Omega$ stands for the anisotropic fluctuations. 
The conventional $1/(4\pi)$ prefactor is introduced in order to recover the background-level results by integrating Eq.~(\ref{eq:omegagw}) over the full solid angle. 

The homogeneous and isotropic quantity $\overbar{\Omega}(\nu)$ at the background level is computed as \cite{Wang:2016ana,Wang:2019kaf} 
\begin{align}\label{eq:omegabar}
%\bar{\Omega}(\nu) = \frac{\nu}{\rho_{c}}\int_{0}^{\eta_{0}} d\eta a(\eta) \int d\theta_{s} \mathcal{R}_{\mathrm{X}}(\theta_{s},t) \frac{dE_{s}}{d\nu_{s}}({\nu}_{s},\theta_{s})\ ,
&\overbar{\Omega}(\nu) = \frac{\nu}{\rho_\mathrm{c}}\int_{0}^{\eta_{0}} \mathrm{d}\eta \mathcal{A}_{X}(\eta,\nu)\ ,\\ 
\label{eq:ax}
&\mathcal{A}_{X}(\eta,\nu)=a(\eta) \int \mathrm{d}\theta_\mathrm{s} \mathcal{R}_{X}(\theta_\mathrm{s},t) \frac{dE_\mathrm{s}}{\mathrm{d}\nu_\mathrm{s}}(\nu_\mathrm{s},\theta_\mathrm{s})\ ,
\end{align}
where the subscript ``$\mathrm{X}$" stands for either \ac{PBH} or \ac{ABH}, the subscript ``$\mathrm{s}$" stands for the source frame.
%the subscripts ``${0}$" and ``$\mathrm{s}$" stand for the observer and source frames, respectively. 
The intrinsic energy spectrum at frequency ${\nu}_\mathrm{s}$ for a given source with parameters $\theta_\mathrm{s}$ is encoded in the function $\mathrm{d}{E}_\mathrm{s}/\mathrm{d}\nu_\mathrm{s}$. In terms of the observed frequency $\nu$, we have ${\nu}_\mathrm{s}=(1 + z)\nu$, where $z$ is the unperturbed redshift. In the frequency domain, $\mathrm{d}E_\mathrm{s}/\mathrm{d}\nu_\mathrm{s}$ is related to the \ac{GW} waveform, for which an \acl{IMR} template with nonprecessing spin correction is used \cite{Ajith:2007kx,Ajith:2009bn}. To be specific, for an individual \ac{BBH} coalescence, it is given by \cite{Zhu:2011bd} 
\begin{equation}
    \frac{dE_s}{d\nu_s} = B \left\{
    \begin{aligned}
        & \nu^{-\frac{1}{3}} \quad\quad\mathrm{for}~\nu<\nu_1\\
        & \nu_1^{-1} \nu^{\frac{2}{3}} \quad\quad\mathrm{for}~\nu_1\leq\nu<\nu_2\\
        & \frac{\nu_1^{-1}\nu_2^{-4/3}\nu^{2}}{\left[1+\left(\frac{\nu-\nu_2}{\sigma/2}\right)^{2}\right]^{2}}   \quad\mathrm{for}~\nu_2\leq\nu<\nu_3
    \end{aligned}
    \right.
\end{equation}
where $B=(G\pi)^{\frac{2}{3}}m_c^{\frac{5}{3}}/3$, and a chirp mass is defined as $m_c=m_1m_2(m_1+m_2)^{-\frac{1}{3}}$ with $(m_1,m_2)$ being two component masses. 
The parameters $(\nu_1,\nu_2,\sigma,\nu_3)$ are given in terms of $(a\eta^2+b\eta+c)/(\pi m_t)$, where $m_t=m_1+m_2$ is the total mass and $\eta=m_1m_2/m_t^2$ is the symmetric mass ratio. The constants $(a,b,c)$ are been given in Table~1 of Ref.~\cite{Ajith:2007kx}. 
Here, a binary inclination angle has been integrated over, and an additional factor is absorbed into $\mathrm{d}{E}_\mathrm{s}/\mathrm{d}\nu_\mathrm{s}$. We assume that the orbits of binaries are circularized due to long evolution. 
%This simplification is viable when one is interested in the high-frequency \aclp{GW}. 
%However, we might consider the eccentricity if we study the lower-frequency detectors such as the \ac{LISA} \cite{Audley:2017drz}. 
This is well justified for the \ac{PBH} binaries formed in the early Universe. 
However, the eccentricity could play an important role in dynamical formation of \ac{ABH} binaries. However, assuming a black hole mass spectrum with power-law index larger than $2$, the LIGO-Virgo Collaboration has excluded \cite{LIGOScientific:2019dag} the merger rates larger than $100~\mathrm{Gpc}^{-3}\mathrm{yr}^{-1}$ for relatively large eccentricities $e>0.1$. On the other hand, the correction to the radiated power of gravitational waves is smaller than $6.7\%$ when $e\leq0.1$ \cite{Maggiore:2007ulw} suggesting that the eccentricity will not significantly alter our theoretical predictions.

The main statistical properties of the \ac{SGWB} anisotropies are encoded in the angular two-point autocorrelation function $\langle \delta\Omega(\nu,\mathbf{e}) \delta\Omega(\nu,\mathbf{e}^\prime) \rangle$, where $\mathbf{e}$ and $\mathbf{e}^\prime$ are two directions with a fixed angular separation. In practice, the modelling is simpler in the harmonic space, where we work in terms of the angular power spectra given by
\begin{align}
\label{eq:aps}
C_{\ell}(\nu)=\frac{2}{\pi}\int \mathrm{d}\ln k ~ {k^{3}} |\delta\Omega_{\ell}(\nu,k)|^{2}\ .
\end{align}
Here the $\delta\Omega_{\ell}$ quantities are line-of-sight integrals over the source functions characterizing all the relevant effects leading to \ac{SGWB} anisotropies. These can be broadly categorized as production and propagation effects, with the former being linked to the inhomogeneous spatial distribution of the \ac{GW} sources, and the latter to the propagation of \acp{GW} in a perturbed Universe.   While we have included all the effects in our analysis, let us stress for clarity that the primary source of anisotropies is rooted in the spatial distribution of the \ac{GW} sources. Instead of showing the complete expression containing all the effects (see Ref.~\cite{Cusin:2018rsq}), here we only present this primary term 
\begin{align}\label{eq:deltal}
\delta\Omega_{\ell}(\nu,k) = \frac{\nu}{4\pi\rho_\mathrm{c}} &\int^{\eta_{0}}_{0} \mathrm{d}\eta \mathcal{A}_{X}(\nu;\eta)\times\nonumber\\
&b_{X}\left(\eta\right)\delta_\mathrm{m}\left(\eta, k\right)(\eta)j_{\ell}(k\Delta\eta)+\cdots\ ,
\end{align}
where $\mathcal{H}$ is the conformal Hubble parameter, $\Delta\eta \equiv \eta_{0}-\eta$ is the look-back time, $j_{\ell}(k\Delta\eta)$ is the spherical Bessel function, $\delta_{m}$ is the dark matter overdensity, and $b_{X}$ is the linear bias function of the population $X$ (either \ac{PBH} or \ac{ABH} in our analysis). In the early Universe formation channel the \ac{PBH} binaries are not expected to be biased with respect to dark matter, which motivates us to consider $b_{\mathrm{PBH}} = 1$. On the other hand, the astrophysical binaries are preferentially formed in larger halos, which are heavily biased with respect to dark matter. We model their bias assuming a simple parametric form $b_{\mathrm{ABH}}=b_1+b_2/D$, where $D$ is the linear growth rate and $b_1 = b_2 = 1$ are chosen as constants \cite{Oguri:2016dgk}. 
%Later on we will also introduce the spectra for relative anisotropies $\widetilde{\delta\Omega}=\delta\Omega(\bar{\Omega}/4\pi)^{-1}$ which can be obtained from the $\delta\Omega$ spectra by a rescaling. 
Cosmological perturbations are obtained by numerically solving the Einstein-Boltzmann equations in the standard model of cosmology. In practice we use the \texttt{CMBquick} package, while adopting the Halofit \cite{Takahashi:2012em} in order to account for nonlinearities of perturbations at smaller scales. 
%One should note that $C_{\ell}$ characterizes the amplitude of the anisotropies in \ac{SGWB} on an angular separation $\sim\pi/\ell$ across the sky. 

Besides modeling the signal, a significant care should be dedicated to the shot-noise estimates. Unlike, for example, the more conventional galaxy number counts, the constituent sources of \ac{SGWB} are not only discretely and randomly distributed in space, but are also discrete in time if the observation time scale is longer than the typical times the binaries spend in a particular frequency band of interest. The discreteness of the spatial distribution of binaries leads to the widely familiar spatial shot noise, while the discreteness of binary mergers in time leads to a temporal or ``popcornlike" shot noise. In the high-frequency regime, e.g. in the \ac{LVK} band, the latter typically dominates over the former \cite{Jenkins:2019uzp,Jenkins:2019nks,Cusin:2019jpv,Canas-Herrera:2019npr,Alonso:2020mva,Mukherjee:2019oma,Bellomo:2021mer}. In contrast, in the low-frequency regimes, e.g. in the \ac{LISA} band \cite{LISA:2017pwj}, \footnote{In this paper we will focus on three experiments --- \ac{LISA} \cite{LISA:2017pwj}, \ac{BBO} \cite{Harry:2006fi} and Ultimate \ac{DECIGO} \cite{Sato:2017dkf}, for the \ac{LISA} band. } the latter is negligible, because the background mainly arises from the inspiraling stage of binaries and thus is almost stationary during the observation window \cite{Canas-Herrera:2019npr,Cusin:2019jhg,Scelfo:2021fqe,Capurri:2021zli}. 
Furthermore, we neglect the contribution from supermassive black hole mergers since they are subdominant with respect to stellar-mass black hole binaries \cite{Erickcek:2006xc}.

The spatial shot noise can be evaluated using the expression \cite{Canas-Herrera:2019npr,Alonso:2020mva}
\begin{align}\label{eq:ssn}
    N_{\ell} = \frac{1}{(4\pi)^{2}} \int_{r_\ast} \frac{\mathrm{d}r}{r^2} \frac{1}{\overbar{n}(r)} \left(\frac{\nu\mathcal{A}_{X}}{\rho_\mathrm{c}}\right)^2 \ ,
\end{align}
where $r = \Delta\eta$ is the comoving distance and $\overbar{n}(r)$ is the comoving number density of binaries at a distance $r$, that emit \acp{GW} in the detection band. For demonstration, we consider a situation where the spectral bandwidth of the detector approximately covers the frequency range from $\nu^\mathrm{low} = 10^{-3}$~Hz to $\nu^\mathrm{high} = 10^{-2}$~Hz. The comoving number density of emitting sources can be estimated as $\overbar{n}(r)=\int_{t(r) + \tau(\nu_s^\mathrm{high})}^{t(r) + \tau(\nu_s^\mathrm{low})}\mathcal{R}_{X}(t^\prime)\mathrm{d}t^\prime$, where $\tau(\nu_\mathrm{s})\simeq 2.18(1.21 M_\odot/m_\mathrm{c})^{5/3}(100\mathrm{Hz}/\nu_\mathrm{s})^{8/3}$ is the coalescence time in units of seconds and $m_\mathrm{c}$ is the source-frame chirp mass, which in our case is given by $m_\mathrm{c}=2^{-1/5}m_0$. We note that while we focus on the \ac{LISA} band in our study, the expressions are in fact generic.

It should be noted that the integral in Eq.~(\ref{eq:ssn}) is divergent for $r_\ast = 0$, and a positive lower limit should therefore be adopted in practice. Effectively, this constitutes in resolving the local sources within $r_\ast$, and subtracting their contribution from the background. In this work we set $r_\ast = 200$ Mpc for concreteness, but our results do not depend strongly on this choice. If we choose a larger $r_\ast$, the shot noise would be smaller. This implies that our choice is a conservative estimate for the LISA band.

\section{Theoretical results}\label{sec:result}

\begin{figure}
    \includegraphics[width =1. \columnwidth]{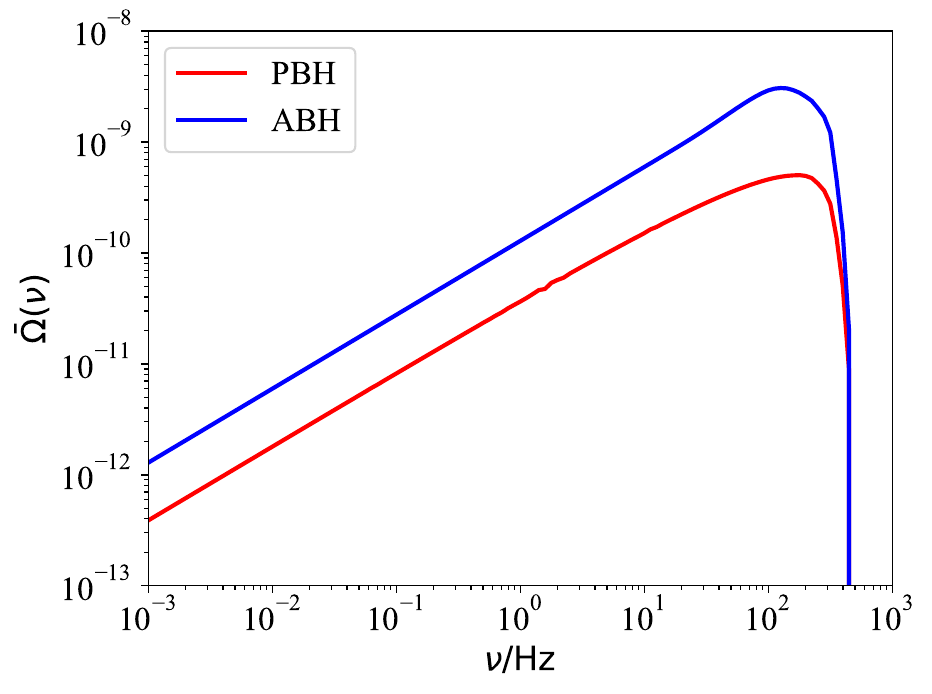}
    \caption{Isotropic component $\bar{\Omega}(\nu)$ in \ac{PBH} (red) and \ac{ABH} (blue) scenarios.}\label{fig:1}
\end{figure}

\begin{figure}
    \includegraphics[width =1. \columnwidth]{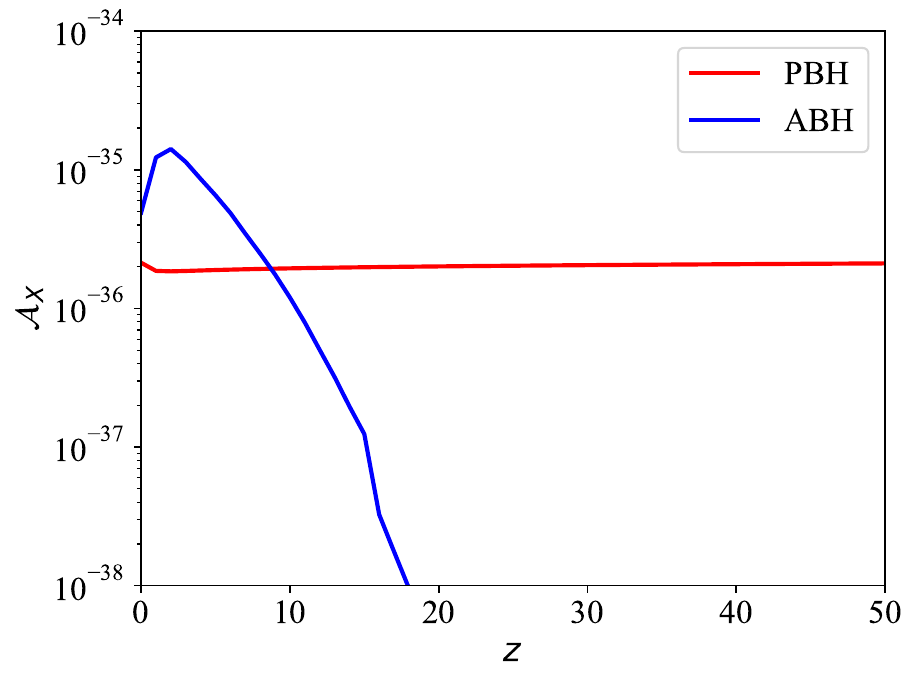}
    \caption{Kernel $\mathcal{A}_{X}(\eta(z),\nu)$ as a function of redshift $z$, at frequency $\nu=10^{-3}$~Hz. The labeling is the same as in Fig.~\ref{fig:1} and is consistent in the rest of the paper.}\label{fig:2}
\end{figure}

Fig.~\ref{fig:1} depicts the isotropic component of the signal, i.e. the monopole $\overbar{\Omega}$ of the \ac{SGWB}, originating from \acs{PBH} (red solid curve) and \acs{ABH} (blue solid curve) binaries. The overall shapes of these two curves are identical due to our assumption of a monochromatic \ac{BH} mass function, while the amplitudes are different. Such differences in spectral amplitudes can be traced back to the difference in merger rates as a function of redshift. This is encoded in Eq.~(\ref{eq:ax}), which we depict as a function of $z$ in Fig.~\ref{fig:2} (the coloring of curves is consistent with that of Fig.~\ref{fig:1}). From these two figures it is clear that the energy density of \ac{SGWB} is mainly contributed by low-redshift \acs{BBH}s. This is an expected result since the \acp{GW} emitted from high-redshift sources are significantly diluted due to the cosmic expansion. This also implies that if the local merger rates of \acp{PBH} and \acp{ABH} are identical, it would be challenging to discriminate the corresponding monopoles \cite{Mukherjee:2021ags}. 
The time dependence of the astrophysical kernel $\mathcal{A}_{\mathrm{PBH}}(\eta)$ is determined in terms of $a(\eta)$, $\mathcal{R}$ and $\mathrm{d}E_s/\mathrm{d}\nu_s$ [see Eq.~(\ref{eq:ax})]. The scale factor is given by $(1+z)^{-1}$, in matter domination the second one scales as $t^{-34/37}\propto (1+z)^{51/37}$. Finally, the redshifted \ac{GW} energy scales as $(1+z)^{-1/3}$. 
Combining we find $(1+z)^{5/111}\simeq(1+z)^{0.045}$, implying an almost flat curve in Fig.~\ref{fig:2}. In contrast, the ABH kernel $\mathcal{A}_{\mathrm{ABH}}$ has a single peak corresponding to the peak in the comoving merger rate in Fig.~\ref{fig:0}. 

\begin{figure}
    \includegraphics[width =1. \columnwidth]{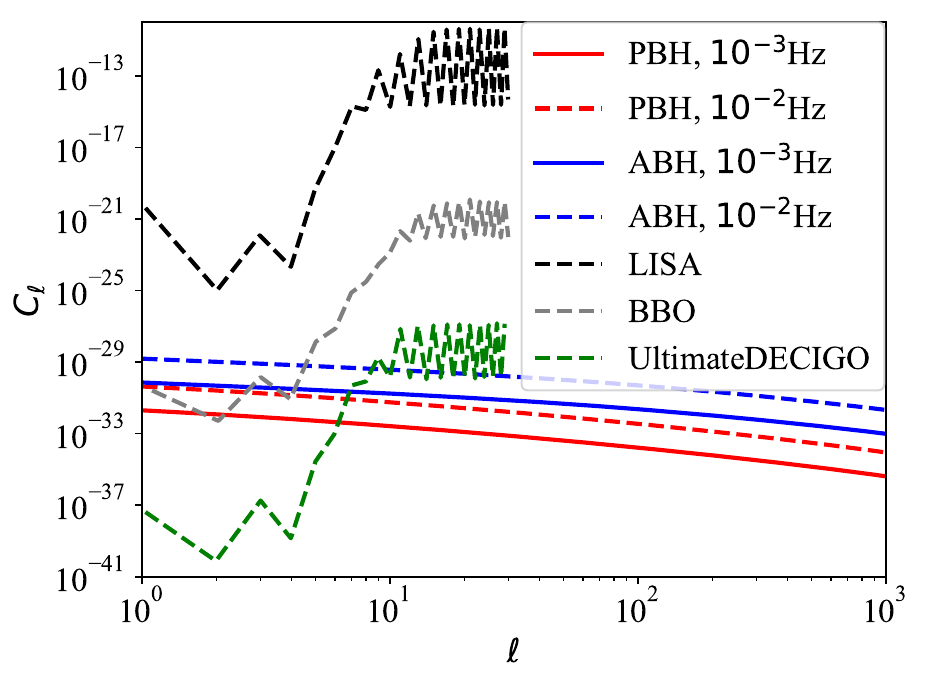}
    \caption{Angular power spectra $C_\ell$ at $\nu=10^{-2}$~Hz (dashed) and $10^{-3}$~Hz (solid). For comparison, the noise power spectra of LISA (black), BBO (gray), and Ultimate DECIGO (green) are plotted for the multipoles $\ell \in [1, 30]$ at frequency $10^{-2}$~Hz (sensitivity curves adopted from \cite{Braglia:2021fxn}).}\label{fig:3}
\end{figure}

After having established the monopole signal, we now move forward to evaluating the angular spectra using Eq.~(\ref{eq:aps}). Fig.~\ref{fig:3} shows the angular power spectra $C_\ell$ for the anisotropies of \ac{SGWB} in \ac{PBH} and \ac{ABH} scenarios. We show the results at two frequencies $10^{-2}$~Hz (dashed curves) and $10^{-3}$~Hz (solid curves) to reveal the frequency dependence of the spectra; see below for further discussion. The two populations seem to have the same dependence for the angular power spectra, even though they have a very different merger rate and kernel. This prediction arises from the fact that the energy density of \ac{SGWB} is mainly contributed by low-redshift sources, since the \acp{GW} emitted from high-redshift sources are significantly diluted. For comparison, we also depict the noise power spectra at multipoles from $\ell = 1$ to $30$, and at $\nu=10^{-2}$~Hz for \ac{LISA} (black dashed curve), \ac{BBO} (gray dashed curve) and Ultimate \ac{DECIGO} (green dashed curve) \cite{Braglia:2021fxn} \footnote{For each experiment, the noise power spectrum has been shown at the peak frequency in Ref.~\cite{Braglia:2021fxn}.  Simply rescaling the noise power spectrum from its peak frequency to 10 mHz, i.e. multiplying it with $[\bar{\Omega}(10\ \mathrm{mHz})/\bar{\Omega}(\mathrm{peak~frequency})]^2$, we obtain a revised noise power spectrum at 10 mHz in Fig.~\ref{fig:3}. This rescaling is reasonable in the sense that the monopole only varies by a few times from the peak frequency to 10 mHz. }. Not surprisingly, such a small signal is beyond the measuring capability of \ac{LISA}, with the predicted signal at $10^{-2}$~Hz being around five orders of magnitude below the expected sensitivity of \ac{LISA}. It should, however, be stressed that detector networks might have a much better sensitivities. 
However, the signal is marginally within the capability of \ac{BBO} for the first four multipoles. In contrast, for the first six multipoles, Ultimate \ac{DECIGO} has the capability to measure the signal, since its expected sensitivity is lower than the predicted signal by eight orders of magnitude. 
%Furthermore, we find from Fig.~\ref{fig:3} that the spectral profiles of $C_\ell$ are similar for \acp{PBH} and \acp{ABH}, except a rescaling of their spectral amplitudes. 

In addition, for a given frequency, e.g., $10^{-3}$~Hz, the spectra of \acp{PBH} and \acp{ABH} share nearly the same profile, while their amplitudes differ. For the sake of a better comparison, it is instructive to consider the spectra of the relative anisotropies $\delta\Omega/\overbar{\Omega}$, instead of the absolute $\delta\Omega$. The corresponding power spectra would assist in evaluating the shape differences, as well as the frequency dependence of the signals.

\begin{figure}
    \includegraphics[width =1. \columnwidth]{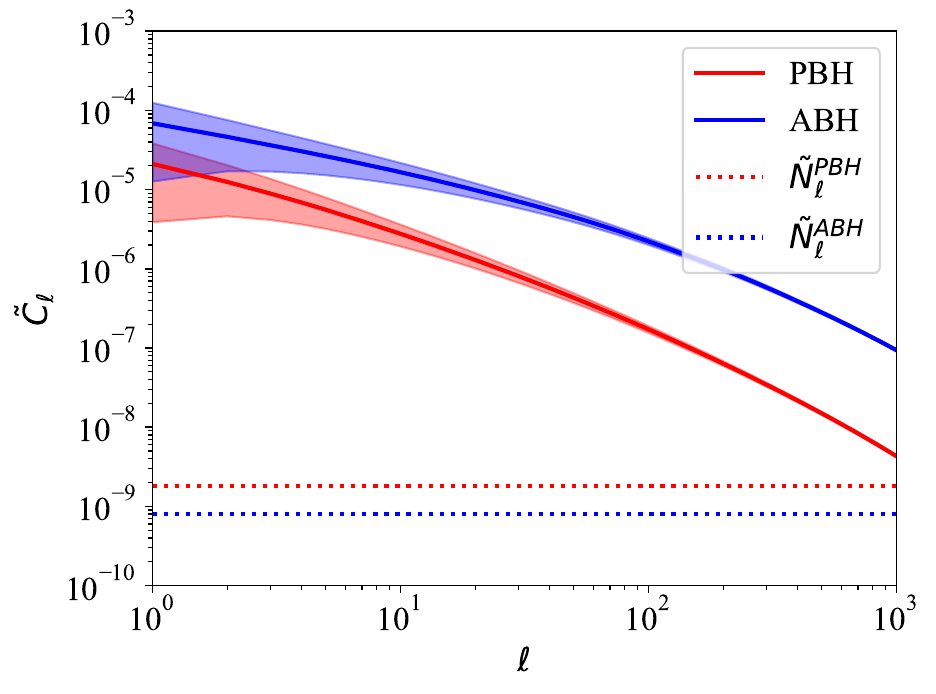}
    \caption{Reduced angular power spectra ${\widetilde{C}_{\ell}}={C_{\ell}} (\overbar{\Omega}/4\pi)^{-2}$. The shaded regions represent the cosmic variance. The reduced shot noise $\widetilde{N}_\ell=N_{\ell}(\overbar{\Omega}/4\pi)^{-2}$ (dotted) is shown for comparison.  }\label{fig:4}
\end{figure}

Indeed, the frequency dependence of $C_{\ell}$ can be reduced if we consider a redefined spectra ${\widetilde{C}_{\ell}}={C_{\ell}} (\overbar{\Omega}/4\pi)^{-2}$, corresponding to the autocorrelations of relative fluctuations  $\delta\Omega(\overbar{\Omega}/4\pi)^{-1}$. 
Our results for such ``reduced'' angular power spectra are shown in Fig.~\ref{fig:4}. By a direct computation we have established that $\widetilde{C}_{\ell}$ does not depend on the frequency. This, in turn, implies that the frequency dependence of $C_\ell$ is completely dominated by the frequency dependence of the monopole $\overbar{\Omega}$. Moreover, the shape differences  between the spectra in \ac{ABH} and \ac{PBH} scenarios are better visible when considering the relative anisotropies, as can be seen in Fig.~\ref{fig:4}. The reduced spectra $\widetilde{C}_{\ell}$ are therefore very useful for identifying the origin of the \ac{SGWB} anisotropies, and do not contain the redundant information present in the $C_{\ell}$ spectra.

While the results in this paper are derived with exact numerical evaluation of the spectra in Eq.~(\ref{eq:aps}), it is useful to consider an approximate treatment, relying on widely used Limber approximation \cite{LoVerde:2008re}. Here, for simplicity, we assume a constant comoving merger rate density, i.e., $\mathcal{R}_{X} = 20 \Gpcyr$, which is compatible with the event rate of $16-61 \Gpcyr$ for \acp{BBH} reported by \ac{LVK} \cite{LIGOScientific:2021psn}. 
Considering $\mathrm{d}E_\mathrm{s}/\mathrm{d}\nu_{s}\sim 10^{53}\mathrm{erg/Hz}$ in the mHz band for \acp{BBH} with component masses of $\simeq10M_\odot$, and using the numerical values $\rho_\mathrm{c}\sim 10^{11}M_\odot/\mathrm{Mpc}^{3}$, $t_0\sim 10\mathrm{Gyr}$ and $M_\odot\sim 10^{54}\mathrm{erg}$, from Eqs.~(\ref{eq:omegabar}) and (\ref{eq:ax}) we get
\begin{align}
\overbar{\Omega} \simeq \frac{\nu t_0 \mathcal{R}_{X}}{\rho_{\mathrm{c}}} \frac{\mathrm{d}E_\mathrm{s}}{\mathrm{d}\nu_\mathrm{s}} \simeq 10^{-13}.  
\end{align}
This result is consistent with our numerical results of $\overbar{\Omega}$ in Fig.~\ref{fig:1} up to an order-one constant prefactor.

Using the Limber approximation \cite{PhysRevD.78.123506}, the angular power spectra $C_\ell$ in Eq.~(\ref{eq:aps}) can be estimated as  \begin{align}\label{eq:limber}
    C_\ell \simeq \frac{1}{(4\pi)^{2}} \left(\ell+\frac{1}{2}\right)^{-1} \int \mathrm{d}k P(k; \eta) \left(\frac{\nu\mathcal{A}_{X}}{\rho_\mathrm{c}}\right)^2 \ ,
\end{align}
where $P(k;\eta)$ denotes the matter power spectrum at scale $k$ and time $\eta$. As consequence of the Limber approximation, the integrand should be understood to be evaluated at look-back time of $\Delta\eta = (\ell + 1/2)/k$. The approximate expression offers an insight into how the kernel $\mathcal{A}_{X}(\eta,\nu)$ affects the spectrum. Indeed, it is clear that since the kernels in \ac{ABH} and \ac{PBH} scenarios are drastically different (see Fig.~\ref{fig:2}) from each other, the resulting angular spectra should also be different (see Fig.~\ref{fig:4}). Additionally, taking into account that the $\mathcal{A}_\mathrm{PBH}$ is a nearly flat function of redshift, and assuming for simplicity that the power spectrum does not depend significantly on redshift either, Eq.~(\ref{eq:limber}) suggests a simple scaling $C_\ell\propto(\ell+1/2)^{-1}$, which is approximately compatible with the numerical behavior seen in Figs.~\ref{fig:3} and \ref{fig:4}. 

We now evaluate the spatial shot noise $N_\ell$ following Eq.~(\ref{eq:ssn}). In order to compare it with the reduced spectra $\widetilde{C}_\ell$, we introduce $\widetilde{N}_\ell = N_{\ell}(\overbar{\Omega}/4\pi)^{-2}$, for which the numerical results (dotted curves) are shown in Fig.~\ref{fig:4}. We conclude that $\widetilde{N}_\ell$ is at the level of $\sim10^{-9}$ at $\nu=10^{-3}$~Hz, and, depending on $\ell$, is smaller than the predicted signal $\widetilde{C}_{\ell}$ by at most $4-5$ orders of magnitude. As a result it can be safely neglected in the \ac{LISA} frequency band and at a vast range of angular scales. Note, however, that the shot-noise becomes more dominant at much smaller scales, corresponding to higher multipoles. 

For the sake of completeness, we additionally present a simple (but crude) estimate for the amplitude of the shot-noise $\widetilde{N}_\ell$. Assuming a constant merger rate, and considering a \ac{BBH} coalescence time $\tau \sim 10^3$ years (roughly spanning the frequency range from $10^{-3}$ Hz to $10^{-2}$ Hz), for the comoving number density of \acp{BBH} we approximately obtain $\overbar{n}\sim\tau\mathcal{R}_{X}\sim10^{5}\mathrm{Gpc}^{-3}$. Eq.~(\ref{eq:ssn}) can be approximated to give
\begin{align}
    N_\ell \sim \frac{1}{\left(4\pi\right)^2}\frac{1}{\overbar{n}t_0}\left(\frac{\nu\mathcal{A}_{X}}{\rho_\mathrm{c}}\right)^2.
\end{align} 
Using $\overbar{\Omega}\sim t_0 \nu\mathcal{A}_{X}/\rho_\mathrm{c}$, we obtain $\widetilde{N}_\ell\sim (t_0^3 \overbar{n})^{-1} \sim 10^{-8}$, which is consistent with the numerical evaluation. 

\begin{figure}
    \includegraphics[width =1. \columnwidth]{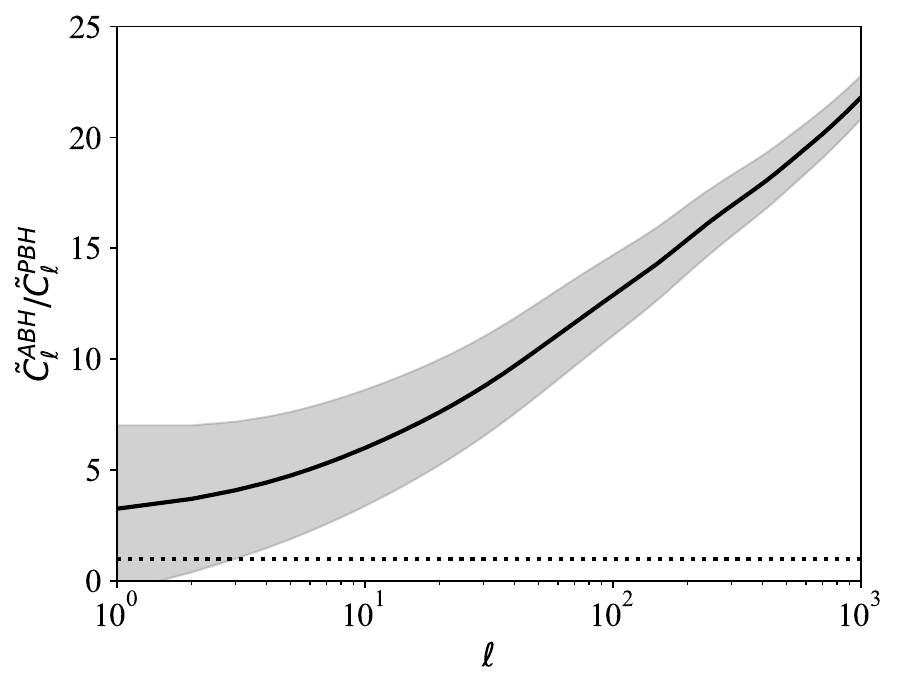}
    \caption{Ratio between the reduced power spectra in \ac{ABH} and \ac{PBH} scenarios (black solid line). Cosmic variance is shown as a gray shaded region. The dotted horizontal line represent the case of identical signals.}\label{fig:5}
\end{figure}

Finally, in order to assess the power of our method, in Fig.~\ref{fig:5} (black solid curve) we show the ratio of the rescaled $\widetilde{C}_{\ell}$ spectra in the \ac{ABH} and \ac{PBH} scenarios. For comparison, we also show the cosmic variance (shaded region), which, for each of the signals, is given by $\sigma(C_\ell)/C_\ell=\sqrt{2/(2\ell+1)}$. For the ratio of two normally-distributed variables $w=u/v$, the variance $\sigma_{w}$ is given by $\sigma_{w}^{2}/w^{2}=\sigma_{u}^{2}/u^{2}+\sigma_{v}^{2}/v^{2}$, where $\sigma_{u}$ and $\sigma_{v}$ denote the variances of $u$ and $v$, respectively. Given the cosmic variances of $u = \widetilde{C}_\ell^{\mathrm{ABH}}$ and $v = \widetilde{C}_\ell^{\mathrm{PBH}}$, we estimate the cosmic variance of the ratio $w=\widetilde{C}_\ell^{\mathrm{ABH}}/\widetilde{C}_\ell^{\mathrm{PBH}}$; the gray-shaded region in Fig.~\ref{fig:5}. The dotted horizontal line represents the case of identical signals, which we aim to rule out. 
%We find that at $\ell<20$ cosmic variance prevents any discrimination between \acp{PBH} and \acp{ABH}. 
%In contrast, it becomes subdominant at $\ell\geq 20$, implying possible discrimination in principle. 
We find that the cosmic variance is overall subdominant with respect to the signal, implying that it would be possible to discriminate the \ac{PBH} scenario from the \ac{ABH} one using the \ac{SGWB} anisotropies. Scaling as approximately a power-law $\ell^{-1/2}$, the cosmic variance is far less dominant at higher multipoles. However, as our result in Fig.~\ref{fig:5} shows, even the lowest multipoles are useful for identifying the \ac{BBH} origin. This is an important conclusion, because the detection of smaller-scale anisotropies is known to be technically challenging. Furthermore, in particular, we expect that with the lowest six multipoles, Ultimate \ac{DECIGO} has the capability to distinguish the predicted signals of different origin, since it is expected to measure the angular power spectrum with high sensitivity, as mentioned before.

\section{Conclusions and discussions}\label{sec:conc}

In this work, we have proposed a novel observational window to probe \acs{PBH} scenario using the anisotropies in the stochastic gravitational wave background. We particularly provided the theoretical modelling of angular correlations and discussed the theoretical observability of the signal in the milli-Hertz frequency band.

We have found $C_\ell\lesssim 10^{-29}$ for the angular power spectra, and $\widetilde{C}_\ell \sim 10^{-9}-10^{-5}$ for the spectra normalized by the isotropic component of the background. We have shown that the shot noise $\widetilde{N}_\ell$ is constant and negligible ($\sim 10^{-12}$) for multipoles $\ell\lesssim\mathcal{O}(10^{3})$. 
While the shot noise could take over the signal at very small angular scales, the latter are not expected to be probed in foreseeable future. Our results demonstrate that cosmic-variance-limited detection of the anisotropies would allow the \ac{ABH} and \ac{PBH} signals to be distinguishable from one another even with poor angular sensitivities. Particularly, cosmic variance scales approximately as $\sim \ell^{-1/2}$, and the detection of correlations with $\ell \lesssim \mathcal{O}(10)$ would already be useful for discriminating the two scenarios from each other (see Fig.~\ref{fig:5}).

As far as the observational prospects are concerned, we have found that the measurement of anisotropies is beyond the capabilities of \ac{LISA}, but marginally (well) within the capabilities of \ac{BBO} (Ultimate \ac{DECIGO}). Particularly, following Ref.~\cite{Bartolo:2022pez}, we have demonstrated that the signal is $\sim 5$ orders of magnitude lower than the \ac{LISA} sensitivity, but is within reach of \ac{BBO} and Ultimate \ac{DECIGO}. Our results, therefore, have interesting observational prospects. Future experimental proposals, as well as improvements in map-making techniques, will provide better sensitivities, therefore better prospects for our results. Additionally, cross-correlations with galaxy distribution are expected to improve the detection prospects as well \cite{Canas-Herrera:2019npr,Alonso:2020mva}.    
%One may wonder if changing the mass function can help in distinguishing better the two signals. In fact, LISA measures $C_\ell$, which is $\tilde{C}_\ell$ multiplied with $\bar{\Omega}^2$. Assuming two different mass functions may slightly change $\tilde{C}_\ell$, because $\tilde{C}_\ell$ is mainly determined by the spatial distribution of low-redshift BBH sources, i.e., in halos or galaxies. Given the existing constraints on the merger rate from Advanced LIGO and Virgo, we find that the value of $\bar{\Omega}$ significantly suppresses $C_\ell$. In fact, based on Fig.~\ref{fig:3}, changing $\bar{\Omega}$ by two orders of magnitude is needed to make ${C}_\ell$ to be observable for LISA. However, only assuming different mass functions could not alter  the orders of magnitude of $\bar{\Omega}$. Therefore, changing the mass functions is expected not to help in distinguishing better the two signals by LISA. Similar discussions are also available to BBO. In contrast, due to high sensitivity, Ultimate DECIGO is expected to distinguish the two signals without the necessity of changing the assumed scenarios, since the difference between the two signals is significant enough, as demonstrated in Fig.~\ref{fig:5}. Changing the mass functions may further strengthen our theoretical expectations of Ultimate DECIGO.  

While demonstrated in the \ac{LISA} frequency band, our results can in principle be generalized to higher-frequency regimes, where the signal could be larger. However, in the \ac{LVK} frequency band the shot-noise component originating from the temporal discreteness of the events is several orders of magnitude larger than the anticipated signal. This time-domain shot-noise is a fundamental problem in the \ac{LVK} band, and no convincing way around it has been proposed so far. There are two promising directions. An interesting approach has been explored in Ref.~\cite{Jenkins:2019nks}, using multiple independent time-segments to estimate the power spectrum. While the approach gives an unbiased estimator of the true \ac{GW} power spectrum, the variance at intermediate to large multipoles dominates over the signal, rendering the approach practically not very useful. Another promising approach, explored in Refs.~\cite{Canas-Herrera:2019npr,Alonso:2020mva} relies on cross-correlating the \ac{GW} anisotropies with galaxy positions. It is still to be shown whether this approach can significantly mitigate the temporal shot-noise. Combining these two approaches may provide a more robust method for mitigating the shot-noise bias.

While the shot-noise is expected to be an important problem, it is useful to note that \ac{LVK} has already presented an upper bound on the power at the lowest multipoles. Following an original description in Ref.~\cite{Thrane:2009fp}, \ac{LVK} inferred an upper bound of $C_\ell\sim 10^{-18}$, for the lowest four multipoles \cite{KAGRA:2021mth}. These limits are at least $\sim 4$ orders of magnitude higher than the expected signal in the corresponding band. This sensitivity could improve significantly in the era of third generation detectors, such as a network of Einstein Telescopes \cite{Mentasti:2020yyd}.

As a final remark let us note that we have made a series of assumptions to simplify our computations in this work. 
First, the assumption of the monochromatic mass distribution of \acp{BH} significantly simplified the numerical computations. Our results would not change significantly when a narrow mass distributions are considered, but broader distributions should be studied separately in a future work. 
%The second assumption said that the eccentricity of binary orbits is negligible for the high-frequency bands, due to a long duration of orbital circularization. 
%However, the eccentricity could be important to the low-frequency bands, which are sensitive to the satellite-borne detectors, e.g. \ac{LISA} \cite{Audley:2017drz}. 
%We leave such a study to future works. 
Second, we have neglected any additional contributions to \ac{SGWB} present, for example, in a number of early Universe models. The \ac{SGWB} from binary mergers could be considered as a foreground for such scenarios. We have also neglected the contributions from late Universe \ac{PBH} binaries since their merger rates are subdominant with respect to the early Universe channel.  Third, in this work we have only compared our theoretical predictions with the sensitivity of a given individual experiment. However, a more detailed analysis involving a network of detectors \cite{Ruan:2020smc,Gong:2021gvw} is required for a better understanding of the practical detectability of our signal. This also should be explored in a future work.

%%%%%%%%%%%%%%%%%%%%%%%%%%%%%%%%%%%%%%%%%%%%%%%

\vspace{2em}
\begin{acknowledgements}
SW is supported by the National Natural Science Foundation of China (Grant No. 12175243), the Key Research Program of the Chinese Academy of Sciences (Grant No. XDPB15) and the science research grants from the China Manned Space Project with No. CMS-CSST-2021-B01. K.K. is supported by KAKENHI Grants No. JP17H01131, No. JP19H05114, No. JP20H04750 and No. JP22H05270. V.V. is supported by the WPI Research Center Initiative, MEXT, Japan and by KAKENHI Grants No. JP20K22348 and No. JP20H04727.
\end{acknowledgements}

\bibliography{sgwb-anisotropy-pbh}

\end{document}